\documentclass[symmetry,article,publish,moreauthors,latex]{mdpi}
\usepackage{amsmath}
\usepackage{amsfonts}
\usepackage{amssymb,bm}
\usepackage{siunitx}
\usepackage{color}
\usepackage{braket}
\usepackage{graphics}

\firstpage{1}
\pubvolume{14}
\issuenum{10}
\articlenumber{2196}
\pubyear{2022}
\copyrightyear{2022}
\datereceived{30 September 2022}
\dateaccepted{14 October 2022}
\datepublished{19 October 2022}
\hreflink{https://doi.org/10.3390/sym14102196} 

\Title{Attractive and repulsive fluctuation-induced pressure in peptide
films deposited on semiconductor substrates}

\TitleCitation{Attractive and repulsive fluctuation-induced pressure in peptide
films deposited on semiconductor substrates}

\Author{
Galina L. Klimchitskaya ${}^{1,2}$,
Vladimir M. Mostepanenko ${}^{1,2,3}$,
Oleg Yu. Tsybin${}^{2}$}

\AuthorNames{
Galina L. Klimchitskaya, Vladimir M. Mostepanenko, Oleg Yu. Tsybin}

\AuthorCitation{
 Klimchitskaya, G.L.; Mostepanenko, V.M.; Tsybin, O.Yu.}

\address{%
${}^{1}$\quad Central Astronomical Observatory at Pulkovo of the Russian Academy of Sciences, Saint Petersburg, 196140, Russia\\
${}^{2}$\quad Peter the Great Saint Petersburg
Polytechnic University, Saint Petersburg, 195251, Russia\\
${}^{3}$\quad Kazan Federal University, Kazan, 420008, Russia}

\corres{Correspondence: vmostepa@gmail.com}

\abstract{We consider the fluctuation-induced (Casimir) pressure in peptide films
deposited on GaAs, Ge, and ZnS substrates which are either in dielectric
or metallic state. Calculations of the Casimir pressure are performed
in the framework of fundamental Lifshitz theory employing the
frequency-dependent dielectric permittivities of all involved materials.
The electric conductivity of semiconductor substrates is taken into
account within the experimentally and thermodynamically consistent
approach. According to our results, the Casimir pressure in peptide
films deposited on dielectric-type semiconductor substrates vanishes
for some definite film thickness, is repulsive for thinner and
attractive for thicker films. The dependence of this effect on the
fraction of water in the film and on the static dielectric permittivity
of semiconductor substrate is determined. For the metallic-type
semiconductor substrates, the Casimir pressure in peptide coatings is
shown to be always repulsive. Possible applications of these results to
the problem of stability of thin coatings in microdevices are discussed.
}


\begin{document}

\newcommand{\ve}{{\varepsilon}}
\newcommand{\kb}{{k_{\bot}}}
\newcommand{\skb}{{k_{\bot}^2}}

\section{Introduction}\label{Intro}

Interaction energies and forces induced between material surfaces
by the zero-point and thermal fluctuations of the electromagnetic
field are the subject of considerable literature (see, e.g., the
monographs \cite{1,2,3,4,5,6,7}). They are known as the van der Waals
forces, Casimir forces, dispersion forces, surface forces etc. Apart
from the fact that the van der Waals and Casimir forces attract much
attention in fundamental physics \cite{1,2,3,4,5,6,7}, during the
last few decades they have found increasing application in
nanotechnology for operation and control of various microdevices
\cite{8,9,10,11,12,13,14,15,16,17,18,19,20,21,22,23}. These
microdevices may contain both inorganic and organic elements, such
as peptides, proteins, and other biological polymers used in organic
electronics \cite{24,25,26,27,28,28a,28b}.

The fluctuation-induced forces and pressures between two plane-parallel
plates kept at any temperature are described by the Lifshitz theory
using the frequency-dependent dielectric permittivities of plate
materials \cite{29,30}. This theory and its generalizations
\cite{31,32,33} are applicable to arbitrarily shaped metallic,
dielectric, and semiconductor bodies. It was also successfully used
to calculate the van der Waals and Casimir forces between organic
surfaces \cite{34,35,36,37}. The formalism of the Lifshitz theory
allows calculation of the fluctuation-induced (Casimir) pressure not
only between two parallel plates, but in a material film either
freestanding in vacuum or deposited on a substrate. This was
illustrated for different inorganic materials of the films and
substrates  \cite{38,39,40,41,42,43}.

The fluctuation-induced free energy and pressure in peptide films,
either freestanding or deposited on dielectric and metallic
substrates, was investigated in  \cite{44,45,46}. It was shown
that for the freestanding peptide films the fluctuation-induced free
energy and pressure are negative and, thus, contribute to the film
stability (the negative pressure is attractive). For peptide films
deposited on metallic substrates both the fluctuation-induced free
energy and pressure are positive (this refers to a repulsion) which
makes peptide coatings on the metallic parts of microdevices less stable.
According to the qualitative estimations \cite{44}, the free energy of
a peptide coating of 100 nm thickness on an Au substrate constitutes
from 5\% to 20\% of the total cohesive energy.

As to peptide films deposited on dielectric (e.g., SiO${}_2$)
substrates, the pressure induced by electromagnetic fluctuations can
be both negative and positive depending on the film thickness and the
fraction of water contained in the film. For the films on a SiO${}_2$
substrate, the fluctuation-induced (Casimir) pressure changes from the
positive (repulsive) to negative (attractive) when the film thickness
increases to above some value. This value varies in the region from
115 to 133~nm depending on the fraction of water in the film \cite{45}.

In this paper, we investigate the fluctuation-induced (Casimir) pressure
in peptide films deposited on semiconductor substrates which are often
used in scientific instruments and for construction of prospective
microdevices. Specifically, the peptide coatings on gallium arsenide
(GaAs), germanium (Ge), and zinc sulphide (ZnS) substrates are
considered. The characteristic feature of doped semiconductors is that
they can be in the dielectric state when the doping concentration is
below some critical value and undergo the Mott-Anderson phase transition
to the metallic state when the doping concentration exceeds this value.
We show that the Casimir pressure in peptide films deposited on
semiconductor substrates strongly depends on whether they are in the
dielectric or metallic state.

It has been known that there is a problem in the Lifshitz theory of
dispersion forces \cite{6,47,47a,48,49}. The theoretical predictions of
the Lifshitz theory for the force acting between dielectric plates
are found to be in disagreement with the measurement data if small
but nonzero conductivity peculiar to all dielectric materials at
any nonzero temperature is taken into account in computations
\cite{50,51,52,53,54,54a}. However, the same theory gives results in
agreement with all measurements if this conductivity is omitted
\cite{6,47,50,51,52,53,54,54a,55}. An important point is also that the
Casimir and Casimir-Polder entropies calculated in the framework
of the Lifshitz theory with included conductivity of dielectric
materials violate the Nernst heat theorem but satisfy it when this
conductivity is disregarded \cite{6,43,47,56,57,58}.

According to our results, the Casimir pressure in peptide films
deposited on semiconductor substrates in the dielectric state
turns into zero for some definite value of the film thickness and
changes from repulsive to attractive for thicker films if the
experimentally and thermodynamically consistent version of the
Lifshitz theory is used. Our computations performed for the peptide
films containing different fractions of water show that the film
thickness resulting in the zero value of the Casimir pressure
increases with decreasing fraction of water in the film and with
increasing static dielectric permittivity of semiconductor
material. It is shown also that the Casimir pressures in peptide
films deposited on semiconductor substrates in the metallic state
is always repulsive. In this case the Casimir pressures in the
films containing different fractions of water differ only slightly.
Possible applications of the obtained results to the problem of
stability of peptide coatings are discussed.

The paper is organized as follows. In Section~2, we briefly present
the formalism of the Lifshitz theory allowing computation of the
Casimir pressure in thin films deposited on thick substrates.
Section~3 presents necessary information about the dielectric
permittivities of substrate materials (GaAs, Ge, and ZnS) and
peptide films containing some fraction of water calculated along
the imaginary frequency axis. Section~4 contains our
computational results for the Casimir pressure in peptide films
deposited on semiconductor substrates as the functions of film
thickness. In Sections~5 and 6, the reader will find  a discussion
 and our conclusions.

\section{Lifshitz formula for the Casimir pressure in a film
deposited on thick substrate}

We consider thick semiconductor substrate, which can be considered as a semispace,
coated by the peptide film of thickness $a$. The dielectric permittivities of
a substrate and a film are denoted as $\ve^{(s)}(\omega)$ and
$\ve^{(p)}(\omega)$, respectively.
In fact a semiconductor substrate in the dielectric state can be viewed as
a semispace if its thickness exceeds $2~\mu$m \cite{59}. In the metallic state,
 semiconductor substrate  can be viewed as a semispace if it is thicker than
 a few hundred nanometers \cite{6}.

In the framework of the Lifshitz theory, the fluctuation-induced pressure in the
film can be treated as occurring in the three layer system consisting of
a substrate semispace, a peptide film, and a vacuum semispace. Assuming that
this system  is at temperature $T$  in thermal equilibrium with the environment,
this pressure is given by \cite{6,41,44,45,46}
\begin{eqnarray}
&&
P(a,T)=-\frac{k_BT}{\pi}\sum_{l=0}^{\infty}{\vphantom{\sum}}^{\prime}
\int_{0}^{\infty}\!\!\kb k^{(p)}(i\xi_l,\kb)d\kb
\nonumber \\
&&
\times\sum_{\alpha}\left[\frac{e^{2a k^{(p)}(i\xi_l,\kb)}}{r_{\alpha}^{(p,v)}(i\xi_l,\kb)
r_{\alpha}^{(p,s)}(i\xi_l,\kb)}-1
\right]^{-1}\!\!\!.
\label{eq1}
\end{eqnarray}

In this equation, $k_B$ is the Boltzmann constant, the prime on the sum in $l$
divides the term with $l=0$ by 2, $\kb$  is the magnitude of the wave vector
projection on the plane of peptide-coated substrate, the sum in
$\alpha=(\rm TM,\,TE)$ is over the transverse magnetic and transverse electric
polarizations of the electromagnetic field, $\xi_l=2\pi k_BTl/\hbar$ are the
Matsubara frequencies, and
\begin{equation}
k^{(p)}(i\xi_l,\kb)=\left[\skb+\ve^{(p)}(i\xi_l)\frac{\xi_l^2}{c^2}
\right]^{1/2}\!\!\!,
\label{eq2}
\end{equation}
\noindent
where the dielectric permittivity of peptide film is calculated at the pure
imaginary frequencies.

The reflection coefficients in  (\ref{eq1}) are defined at the peptide-vacuum
boundary plane
\begin{eqnarray}
&&
r_{\rm TM}^{(p,v)}(i\xi_l,\kb)=\frac{k^{(p)}(i\xi_l,\kb)-\ve^{(p)}(i\xi_l)
q (i\xi_l,\kb)}{k^{(p)}(i\xi_l,\kb)+\ve^{(p)}(i\xi_l)q (i\xi_l,\kb)},
\nonumber \\
&&
r_{\rm TE}^{(p,v)}(i\xi_l,\kb)=\frac{k^{(p)}(i\xi_l,\kb)-
q (i\xi_l,\kb)}{k^{(p)}(i\xi_l,\kb)+q (i\xi_l,\kb)},
\label{eq3}
\end{eqnarray}
\noindent
and at the peptide-substrate boundary plane
\begin{eqnarray}
&&\hspace*{-5mm}
r_{\rm TM}^{(p,s)}(i\xi_l,\kb)=\frac{\ve^{(s)}(i\xi_l)k^{(p)}(i\xi_l,\kb)-
\ve^{(p)}(i\xi_l)k^{(s)}(i\xi_l,\kb)}{\ve^{(s)}(i\xi_l)k^{(p)}(i\xi_l,\kb)+
\ve^{(p)}(i\xi_l)k^{(s)}(i\xi_l,\kb)},
\nonumber \\
&&\hspace*{-5mm}
r_{\rm TE}^{(p,s)}(i\xi_l,\kb)=\frac{k^{(p)}(i\xi_l,\kb)-
k^{(s)}(i\xi_l,\kb)}{k^{(p)}(i\xi_l,\kb)+k^{(s)}(i\xi_l,\kb)},
\label{eq4}
\end{eqnarray}
\noindent
where
\begin{eqnarray}
&&
q(i\xi_l,\kb)=\left(\skb+\frac{\xi_l^2}{c^2}\right)^{1/2}\!\!\!,
\nonumber \\
&&
k^{(s)}(i\xi_l,\kb)=\left[\skb+\ve^{(s)}(i\xi_l)\frac{\xi_l^2}{c^2}
\right]^{1/2}\!\!\!.
\label{eq5}
\end{eqnarray}

Equations (\ref{eq1})--(\ref{eq5}) allow computations of the fluctuation-induced
Casimir pressure in peptide films deposited on a semiconductor substrate if
the required information about the values of film and substrate
permittivities at sufficiently large number of Matsubara frequencies
is available. This
information is collected in the next section.

\section{Dielectric functions of different semiconductors and peptide along
the imaginary frequency axis}

We begin with the gallium arsenide. The measured optical data for the real and imaginary
parts of the complex index of refraction for this semiconductor are presented in
 \cite{60} over the wide frequency region from $\hbar\omega_{\min}=0.00124~$eV
to $\hbar\omega_{\max}=155~$eV. Based on these data, we obtain the imaginary part
of GaAs dielectric permittivity
${\rm Im}\,\ve^{(s)}(\omega)= 2{\rm Re}\,n(\omega){\rm Im}\,n(\omega)$.
In the dielectric state, it is extrapolated to the frequency region
$0\leqslant\omega\leqslant\omega_{\min}$ by ${\rm Im}\,\ve^{(s)}(\omega)=0$.
Then, the dielectric permittivity at the pure imaginary Matsubara frequencies,
$\ve(i\xi_l)$, was computed by the standard procedure \cite{6} using
the Kramers-Kronig relation
\begin{equation}
\ve^{(s)}(i\xi_l)=1+\frac{2}{\pi}\int_{0}^{\infty}
\frac{\omega{\rm Im}\,\ve^{(s)}(\omega)}{\omega^2+\xi_l^2}
d\omega.
\label{eq6}
\end{equation}
\noindent
In doing so an extrapolation of the optical data to the region of high frequencies,
$\omega>\omega_{\max}$, is not needed.

\begin{figure}[!b]
\vspace*{-4.7cm}
\centerline{\hspace*{0cm}
\includegraphics[width=3.5in]{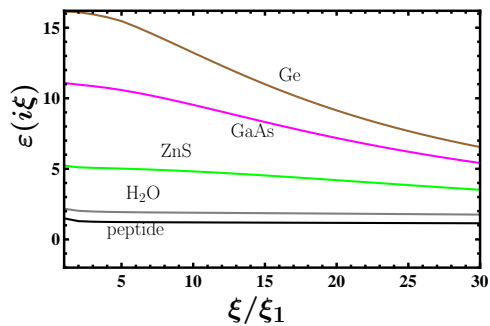}}
\vspace*{-2.5cm}
\caption{\label{fg1}
The dielectric permittivities of germanium, gallium arsenide,
zinc sulphide, water, and pure peptide along the imaginary frequency
axis are shown as the functions of imaginary frequency normalized to
the first Matsubara frequency in the region $\xi > \xi_1$. }
\end{figure}

In Figure~\ref{fg1}, the obtained behavior of the dielectric permittivity of GaAs
along the imaginary frequency axis is shown as the function of $\xi/\xi_1\geqslant 1$.
The value of the static permittivity of GaAs is $\ve^{(s)}(0)=13.0$ \cite{60}.

Next, we consider the dielectric permittivity of germanium. The optical data for
${\rm Re}\,n(\omega)$ and ${\rm Im}\,n(\omega)$ of Ge are also contained in \cite{60}.
They are presented in the region from $\hbar\omega_{\min}=0.00248~$eV
to $\hbar\omega_{\max}=525~$eV. Using these data, the imaginary part of
$\ve^{(s)}(\omega)$ was obtained in the same way as for GaAs. The dielectric
permittivity of Ge at the imaginary Matsubara frequencies was found by the
Kramers-Kronig relation (\ref{eq6}).

The resulting behavior of the dielectric permittivity of Ge
along the imaginary frequency axis is shown as the function of $\xi/\xi_1$
by the top line in Figure~\ref{fg1}.
The static permittivity of Ge in the dielectric state is
$\ve^{(s)}(0)\approx16.2$ \cite{60}.

The last substrate semiconductor considered here is zinc sulphide. By contrast to Ge,
the dielectric permittivity of ZnS with cubic crystal lattice admits sufficiently
exact analytic representation along the imaginary frequency axis (the available
analytic expressions for Ge are only within rather narrow intervals along the real
frequency axis \cite{61}). It is given by the contributions of
ultraviolet and infrared frequencies \cite{62}
\begin{equation}
\ve^{(s)}(i\xi_l)=1+\frac{C_{\rm IR}\omega_{\rm IR}^2}{\omega_{\rm IR}^2+\xi_l^2}
+\frac{C_{\rm UV}\omega_{\rm UV}^2}{\omega_{\rm UV}^2+\xi_l^2},
\label{eq7}
\end{equation}
\noindent
where $C_{\rm IR}=3.27$, $C_{\rm UV}=4.081$,
$\omega_{\rm IR}=5.4\times 10^{13}~$rad/s, and
$\omega_{\rm UV}=9.39\times 10^{15}~$rad/s.
At zero Matsubara frequency one obtains $\ve^{(s)}(0)=8.35$.

In Figure~\ref{fg1}, the behavior of the dielectric permittivity of ZnS
along the imaginary frequency axis is shown as the function of $\xi/\xi_1$.

In the above, we considered the dielectric properties of semiconductors in the
dielectric state. However, as mentioned in Section~I, at any nonzero temperature
dielectric materials are characterized by some nonzero electric conductivity
$\sigma(\omega,T)$. The contribution of this conductivity to the dielectric
permittivity can be taken into account by means of the Drude-like term.
In doing so the values of dielectric permittivity
of semiconductor substrate at the Matsubara frequencies with account of conductivity
are given by
\begin{equation}
 \tilde{\ve}^{(s)}(i\xi_l)={\ve}^{(s)}(i\xi_l)+
 \frac{4\pi\sigma(i\xi_l,T)}{\xi_l},
\label{eq8}
\end{equation}
\noindent
where for dielectric-type semiconductors $\sigma(i\xi_l,T)$ goes to zero
exponentially fast with vanishing temperature.

In Section~1 it was mentioned also that with increasing doping concentration
the semiconductors under consideration here undergo the phase transition
to a metallic state. For metallic-type semiconductors, the total dielectric
permittivity at the Matsubara frequencies can be again represented by
 (\ref{eq8}) but with important distinctive property that the
conductivity  $\sigma(i\xi_l,T)$ may be by several orders of magnitude larger
and it does not go to zero when the temperature vanishes.

In Section~4, we show that the account of nonzero electric conductivity of a
semiconductor substrate can have  a pronounced effect on the fluctuation-induced
Casimir pressure in peptide coating and requires special care for reaching a
physically plausible insight.

Now we turn our attention to the dielectric permittivity of peptide film.
This subject involves difficulties because there are different kinds of
peptides and for none of them the optical data were examined over a sufficiently
wide frequency range. In  \cite{44} the electrically neutral 18-residue
zinc finger peptide was chosen as a basic sample. The imaginary part of the
dielectric permittivity of this peptide in the microwave region was
investigated in  \cite{63}. It was found that $\ve^{(p)}(0)=15$.
To estimate the dielectric properties of peptide film in the ultraviolet and
infrared regions, the data computed for a cyclic tripeptide RGD-4C in
 \cite{64} have been used \cite{44} based on the fact that the films
formed by this peptide and zinc finger peptide are rather similar.

As a result, the dielectric permittivity of our model peptide as the function
of imaginary frequency normalized to the first Matsubara frequency is shown
in Figure~\ref{fg1} by the bottom line in the interval $\xi/\xi_1\geqslant 1$.

It should be noted that peptide films usually contain some fraction of water
which plays the role of a plasticizer providing the required functional
properties of a film \cite{65}. Below we compute the Casimir pressure in
peptide coatings consisting both of pure peptide with the dielectric
permittivity $\ve^{(p)}(i\xi_l)$ and containing the volume fraction $\Phi$ of
water. The dielectric permittivity of such films, $\ve_{\Phi}^{(p)}(i\xi_l)$,
can be found from the mixing formula suitable for the molecules of irregular
shape \cite{66}
\begin{equation}
  \frac{\ve_{\Phi}^{(p)}(i\xi_l)-1}{\varepsilon_{\Phi}^{(p)}(i\xi_l)+2}
=
  \Phi\frac{\varepsilon^{(w)}(i\xi_l)-1}{\varepsilon^{(w)}(i\xi_l)+2}+
  (1-\Phi)\frac{\varepsilon^{(p)}(i\xi_l)-1}{\varepsilon^{(p)}(i\xi_l)+2},
\label{eq9}
\end{equation}
\noindent
which follows from the Clausius-Mossotti equation.

Here, the dielectric permittivity of water at the Matsubara frequencies
can be presented in the form \cite{62}
\begin{equation}
  \varepsilon^{(w)}(i\xi_l)=1+\frac{B}{1+\tau\xi_l}+
  \sum_{j=1}^{11}\frac{C_j{\omega^2_j}}{{\omega^2_j}+
  {\xi_l}^2+{g_j\xi_l}},
\label{eq10}
\end{equation}
\noindent
where the second (Debye) term describes the orientation of permanent dipoles,
the oscillator terms with $j=1,\,2,\,\ldots,\, 5$ correspond to the infrared region,
and the oscillator terms with $j=6,\,7,\,\ldots,\, 11$ present the contribution of
ultraviolet frequencies [see  \cite{44,62}  for the numerical values of all
parameters entering  (\ref{eq10})].

In Figure~\ref{fg1}, the dielectric permittivity of water computed by  ~(\ref{eq10})
is shown as the function of $\xi/\xi_1\geqslant 1$. At zero Matsubara frequency it holds
\begin{equation}
  \varepsilon^{(w)}(0)=1+{B}+
  \sum_{j=1}^{11}{C_j}\approx 81.2.
\label{eq11}
\end{equation}

In the next section, the obtained information about the dielectric properties of
peptide films and semiconductor substrates is used to compute the
fluctuation-induced Casimir pressure.

\section{Computational results for the Casimir pressure in peptide films}

All computations below were performed using the Lifshitz formula (\ref{eq1})
at room temperature $T=300~$K with
the reflection coefficients (\ref{eq3}) and (\ref{eq4}). The semiconductor substrates in both dielectric and metallic
states are considered. We deal with the peptide films containing
$\Phi=0$\%, 10\%, 25\%, and 40\% fractions of water. The computational
results  are presented for GaAs, Ge, and ZnS substrate in the following
subsections.

\subsection{Gallium arsenide substrate}

We computed the Casimir pressure (\ref{eq1}) substituting there the
dielectric permittivity of GaAs, $\ve^{(s)}(i\xi_l)$, in the dielectric
state shown in Figure~\ref{fg1}, the dielectric permittivity of pure
peptide, $\ve^{(p)}(i\xi_l)$, ($\Phi=0$), also shown in Figure~\ref{fg1},
and the dielectric permittivities of peptide films,
$\ve_{\Phi}^{(p)}(i\xi_l)$, with $\Phi=0.1$, 0.25, and 0.4 found by using
 (\ref{eq9}). At $\xi_0=0$  (\ref{eq9}) results in
$\ve_{0.1}^{(p)}(0)=16.5$,
$\ve_{0.25}^{(p)}(0)=19.2$, and $\ve_{0.4}^{(p)}(0)=22.9$.

\begin{figure}[!b]
\vspace*{-3.3cm}
\centerline{\hspace*{.0cm}
\includegraphics[width=3.8in]{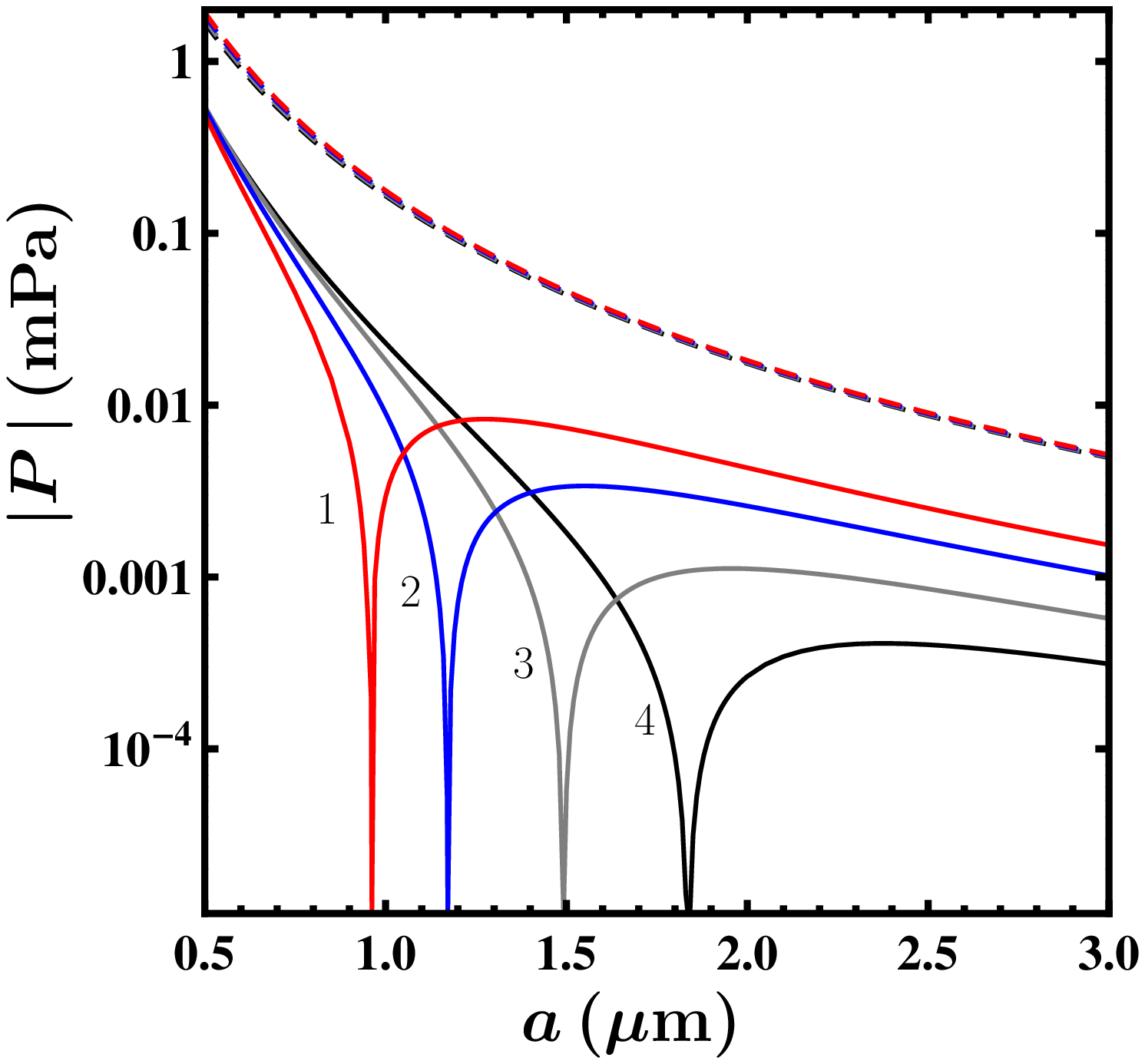}}
\vspace*{-3.1cm}
\caption{\label{fg2}
The magnitudes of the Casimir pressure in peptide films
containing $\Phi$= 0.4, 0.25, 0.1, and 0 fractions of water
deposited on a GaAs substrate in the dielectric state are shown as
the functions of film thickness by the solid lines labeled 1, 2, 3,
and 4, respectively. For a GaAs substrate in the metallic state the
pressure magnitudes in the same peptide films are shown by the
dashed lines (see the text for futher discussion). }
\end{figure}
The magnitudes of the obtained Casimir pressures in peptide films with
$\Phi=0.4$, 0.25, 0.1, and 0 fractions of water on a GaAs substrate are
shown in Figure~\ref{fg2} as the functions of film thickness in the logarithmic
scale by the lines labeled 1, 2, 3, and 4, respectively.

As is seen in Figure~\ref{fg2}, for all four peptide films with different
fractions of water the Casimir pressure turns into zero for some definite
film thickness. For the films with 0.4, 0.25, 0.1, and 0 fractions of water,
this happens for the film thicknesses 0.957, 1.172, 1.492, and $1.84~\mu$m,
respectively. For thinner films, the Casimir pressure is positive, i.e.,
repulsive, whereas for thicker films the pressure is negative, i.e.,
attractive.
Depending on the relationship between the permittivities of
a dielectric substrate and dielectric coating along the imaginary frequency
axis, similar effect takes place for inorganic films possessing a crystalline
structure (for instance, for Al${}_2$O${}_3$ coatings on a high-resistivity Si
substrate) \cite{43}.
One can conclude that to be stable the peptide coating on
GaAs substrate should be sufficiently thick. According to Figure~\ref{fg2},
for peptide films containing smaller fraction of water the film stability
is reached for thicker films. Note that for pure water films
without peptide the Casimir pressure vanishes for much smaller film
thicknesses.

Now we consider the impact of nonzero conductivity of GaAs   substrate on
the Casimir pressure in a peptide film. The major contribution to this impact
is given by the values of reflection coefficient $r_{\rm TM}^{(p,s)}$
defined in  (\ref{eq4}) at zero Matsubara frequency.
Calculating the reflection coefficients (\ref{eq4}) at $\xi_0=0$ ignoring
the electric conductivity, i.e., with the finite value of semiconductor
dielectric permittivity at zero frequency $\ve^{(s)}(0)$
(see in Section~3),
one obtains
\begin{eqnarray}
&&
r_{\rm TM}^{(p,s)}(0,\kb)=
\frac{\ve^{(s)}(0)-\ve^{(p)}(0)}{\ve^{(s)}(0)+\ve^{(p)}(0)},
\nonumber \\
&&
r_{\rm TM}^{(p,s)}(0,\kb)=0.
\label{eq12}
\end{eqnarray}

If, however, the electric conductivity $\sigma$ of a semiconductor substrate
is taken into account, i.e., we calculate the reflection coefficients (\ref{eq4})  substituting the dielectric permittivity $\tilde{\ve}^{(s)}(i\xi_l)$ defined in
 (\ref{eq8}) in place of ${\ve}^{(s)}(i\xi_l)$, the result is
\begin{equation}
r_{\rm TM}^{(p,s)}(0,\kb)=1,
\qquad
r_{\rm TM}^{(p,s)}(0,\kb)=0.
\label{eq13}
\end{equation}

We emphasize that the difference between the values of $r_{\rm TM}^{(p,s)}(0,\kb)$
in  (\ref{eq12}) and (\ref{eq13}) does not depend on the value of $\sigma$,
but only on the presence of the second term on the right-hand side of
 (\ref{eq8}), i.e., it is the same whether we deal with metallic- or
dielectric-type semiconductors with taken into account electric conductivity.
It is significant also that when computing the Casimir pressure in peptide
coating by the Lifshitz formula (\ref{eq1}) using either the dielectric
permittivity ${\ve}^{(s)}(i\xi_l)$ or $\tilde{\ve}^{(s)}(i\xi_l)$ of the
semiconductor substrate, the major difference is obtained  only due to the
different values of the reflection coefficient $r_{\rm TM}^{(p,s)}(0,\kb)$
in  (\ref{eq12}) and (\ref{eq13}), whereas the contributions of all terms
with $l\geqslant 1$ are almost the same.

In Figure~\ref{fg2}, the computational results for the Casimir pressures in
peptide films obtained with account of the substrate conductivity by means
of  (\ref{eq8}) are shown as the functions of film thickness by the
four dashed lines counted from top to bottom for the peptide films
containing $\Phi=0.4$, 0.25, 0.1, and 0 fractions of water, respectively.
As is seen in Figure~\ref{fg2}, the four dashes lines almost overlap.
This means that when the electric conductivity of a substrate is taken into
account the fraction of water in the peptide film makes only a minor impact
on the fluctuation-induced pressure.

According to the discussion presented in Section~1, for the dielectric-type GaAs
substrate the experimentally and thermodynamically consistent results for
the Casimir pressure in peptide coatings are given by the solid lines 1--4 in
Figure~\ref{fg2}. As to the GaAs substrate in a metallic state, the Casimir
pressures in peptide coatings in this case are given by the dashed lines
in Figure~\ref{fg2}.

The physical explanation why the seemingly minor difference between  the values
of the reflection coefficient $r_{\rm TM}^{(p,s)}(0,\kb)$
in  (\ref{eq12}) and (\ref{eq13}) leads to so big difference between
the solid and dashed lines in Figure~\ref{fg2} is the following.
In the configuration of a peptide film deposited on a semiconductor substrate
the contribution of all Matsubara terms with $l\geqslant 1$ in the Lifshitz
formula (\ref{eq1}) is always positive and, thus, leads to the repulsion
for both dielectric- and metallic-type semiconductors.

As to the contribution of the term with zero Matsubara frequency in
 (\ref{eq1}), it is also positive for a metallic-type semiconductors
but may be both positive and negative (i.e., attractive) for
dielectric-type semiconductor. Really, according to  (\ref{eq12}) the
sign of the reflection coefficient $r_{\rm TM}^{(p,s)}(0,\kb)$ depends
on the relationship between $\ve^{(s)}(0)$ and $\ve^{(p)}(0)$.
For a GaAs substrate and peptide films with any fraction of water $\Phi$
from 0 to 0.4 it holds $\ve^{(s)}(0)<\ve^{(p)}(0)$, i.e.,
$r_{\rm TM}^{(p,s)}(0,\kb)<0$. Taking into account that according to
 (\ref{eq3})
\begin{equation}
r_{\rm TM}^{(p,v)}(0,\kb)=\frac{1-\ve^{(p)}(0)}{1+\ve^{(p)}(0)}<0
\label{eq14}
\end{equation}
\noindent
and that at $\xi_0=0$ the TE polarization does not contribute to the
Casimir pressure, one finds that the term of the Lifshitz formula
(\ref{eq1}) with $l=0$ is negative, i.e., produces an attraction.
Considering that with an increase of film thickness the relative role
of the zero-frequency term in the total pressure increases, this leads
to the change of repulsion with attraction as is demonstrated by the
solid lines in Figure~\ref{fg2}. Another situation is illustrated by the
Ge substrate and peptide film which does not contain water (see below).

\subsection{Germanium substrate}

For peptide films deposited on a Ge substrate in the dielectric
state, computations of the Casimir pressure were performed in the
same way as described above for the case of GaAs substrate. In doing
so the dielectric permittivity of Ge along the imaginary frequency
axis was used (see the top line in Figure~\ref{fg1}).

\begin{figure}[!b]
\vspace*{-3.3cm}
\centerline{\hspace*{.0cm}
\includegraphics[width=3.8in]{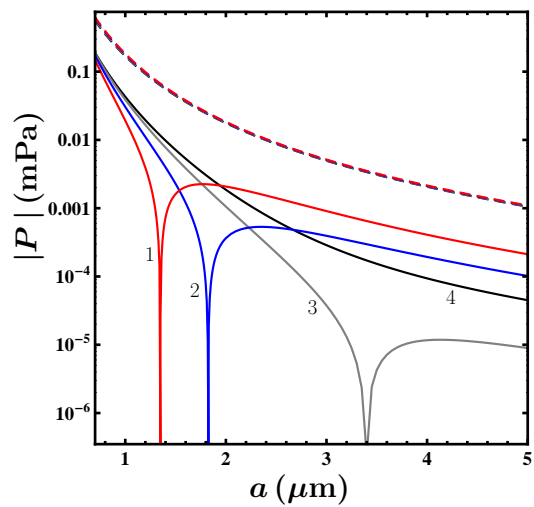}}
\vspace*{-3.1cm}
\caption{\label{fg3}
The magnitudes of the Casimir pressure in peptide films
containing $\Phi$= 0.4, 0.25, 0.1, and 0 fractions of water
deposited on a Ge substrate in the dielectric state are shown as
the functions of film thickness by the solid lines labeled 1, 2, 3,
and 4, respectively. For a Ge substrate in the metallic state the
pressure magnitudes in the same peptide films are shown by the
overlapping dashed lines. }
\end{figure}

In Figure~\ref{fg3}, the computational results for the peptide films
with $\Phi= 0.4$, 0.25, 0.1, and 0 fractions of water deposited on a
Ge substrate are shown as the functions of film thickness in the
logarithmic scale by the solid lines labeled 1, 2, 3, and 4,
respectively.

{}From Figure~\ref{fg3} it is seen that for a Ge substrate in the
dielectric state the Casimir pressure in peptide films containing
$\Phi= 0.4$, 0.25, and 0.1 fractions of water turnes into zero
for the film thicknesses equal to 1.349, 1.824, and 3.359 $\mu$m,
respectively. For thinner films, the fluctuation-induced pressure
in the film is repulsive whereas for thicker films it is attractive
making the film more stable.

The change of the pressure sign is explained in the same way as
was considered in section~4.1 in the case of GaAs substrate. Thus, for
the peptide films containing nonzero fractions of water $\Phi$ under
consideration here it holds
$\ve^{(s)}(0) = 16.2 < \ve^{(p)}_{\Phi}(0)$. As a result,
the zero-frequency term in the Lifshitz formula is negative, i.e.,
contributes to the attraction. With increasing film thickness, the
relative role of this term in the total pressure increases. Because
of this, for sufficiently thick films the total Casimir pressure
becomes attractive.

An alternative situation takes place for a pure peptide film
($\Phi = 0$) deposited on a Ge substrate. In this case,
$\ve^{(s)}(0)= 16.2 > \ve^{(p)}(0)$ = 15.0. As a
result, the reflection coefficient $r^{(p,s)}(0,k_{\bot})$ in
 (\ref{eq12}) becomes positive. Taking into account  ~(\ref{eq14}),
this leads to the positive zero-frequency contribution to the Lifshitz
formula (\ref{eq1}), i.e., the Casimir pressure in the pure peptide
film deposited on a Ge substrate remains repulsive for films of
any thickness. This is illustrated by the solid line labeled 4 in
Figure~\ref{fg3}.

For the metallic-type Ge substrate described by the dielectric
permittivity (\ref{eq8}), the Casimir pressure in peptide films with
any fraction of water is repulsive. In this case, the computational
results are shown by the dashed lines in Figure~\ref{fg3} which almost
overlap for different fractions of water $\Phi$ in a film. Here, the
difference in the values of the Casimir pressure with the case of
a GaAs substrate shown in Figure~\ref{fg2} is only quantitative.

\subsection{Zinc sulfide substrate}

The Casimir pressure in peptide films deposited on a ZnS substrate
is characterized by typically the same behavior as for the case of
GaAs substrate but the change from repulsive to attractive forces
takes place for thinner films. All computations are performed by the
Lifshitz formula (\ref{eq1}) as described above. However, for the
dielectric permittivity of ZnS at the Matsubara frequencies, in place
of the optical data \cite{60} and dispersion relation (\ref{eq6}),
it is now possible to employ the analytic representation (\ref{eq7}).

\begin{figure}[!t]
\vspace*{-3.3cm}
\centerline{\hspace*{.0cm}
\includegraphics[width=3.8in]{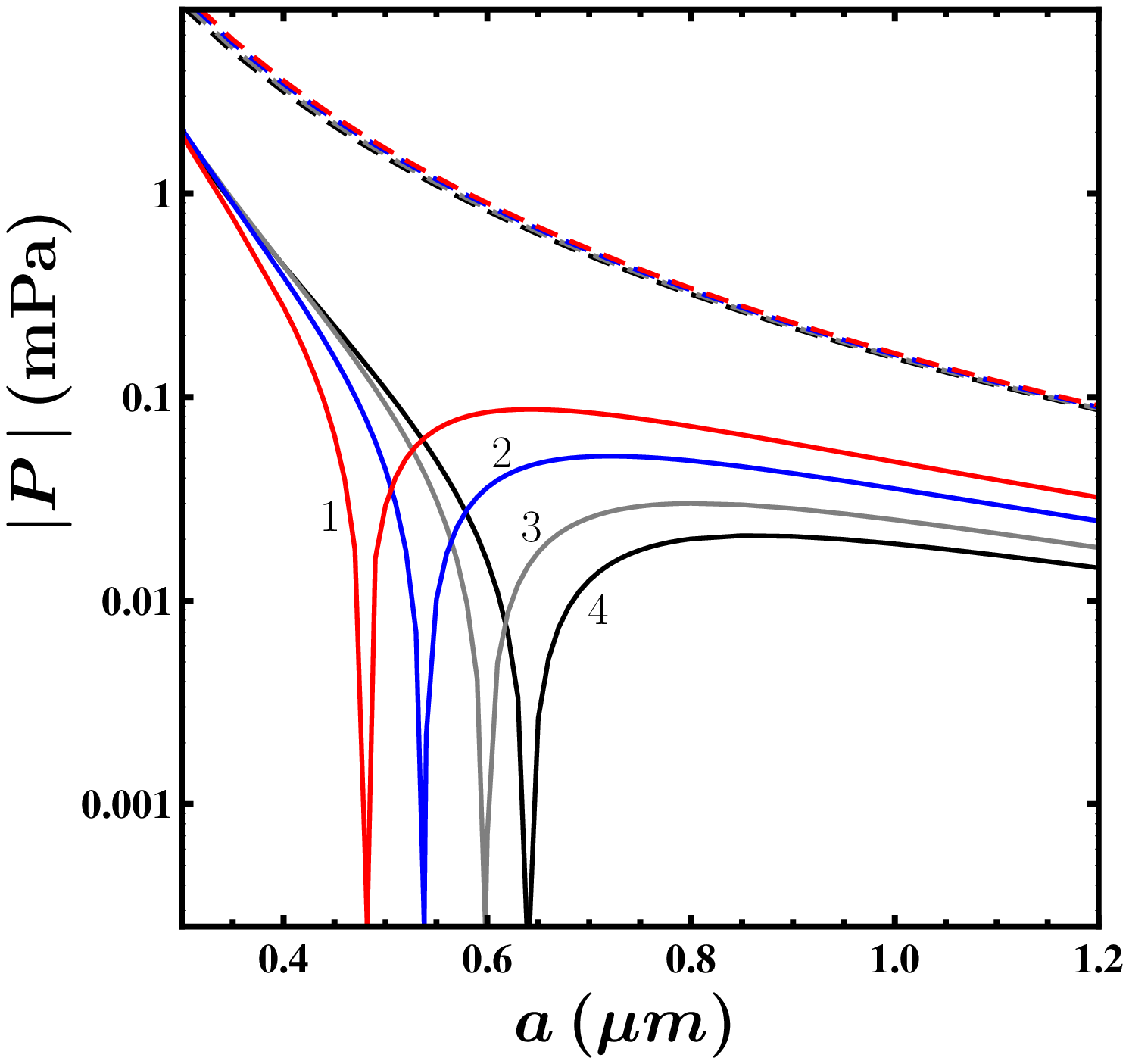}}
\vspace*{-3.1cm}
\caption{\label{fg4}
The magnitudes of the Casimir pressure in peptide films
containing $\Phi$= 0.4, 0.25, 0.1, and 0 fractions of water
deposited on a ZnS substrate in the dielectric state are shown as
the functions of film thickness by the solid lines labeled 1, 2, 3,
and 4, respectively. For a ZnS substrate in the metallic state the
pressure magnitudes in the same peptide films are shown by the
dashed lines (see the text for futher discussion). }
\end{figure}

The computational results for the Casimir pressure in peptide
films with $\Phi= 0.4$, 0.25, 0.1, and 0 fractions of water
deposited on a ZnS substrate in the dielectric state are shown in
Figure~\ref{fg4} as the functions of film thickness in the logarithmic
scale by the four solid lines labeled 1, 2, 3, and 4, respectively.

As is seen in Figure~\ref{fg4}, for the films containing
$\Phi= 0.4$, 0.25, 0.1, and 0 fractions of water the Casimir pressure
vanishes for the film thicknesses of 0.479, 0.537, 0.598, and
0.642~$\mu$m, respectively. Similar to the cases of GaAs and Ge
substrates, the Casimir pressures for thinner films are repulsive
and for thicker films -- attractive. The latter makes the peptide
coatings more stable. In analogy to other substrate semiconductors
considered above, for peptide films containing less water the Casimir
pressure changes its sign for thicker films. The eventual reason
why for ZnS substrate the change of sign of the pressure in the film
containing some fixed fraction of water occurs for smaller film
thickness than for GaAs and Ge substrates is that the static
dielectric permittivity of ZnS, $\epsilon^{(s)}(0)= 8.35$, is
smaller than for GaAs and Ge (13.0 and 16.2, respectively).

For a ZnS substrate in the metallic state the dielectric permittivity
at the pure imaginary Matsubara frequencies is given by  (\ref{eq8})
where $\epsilon^{(s)}(i\xi_l)$ is again expressed by  (\ref{eq7}).
In this case, the Casimir pressures in peptide films containing
$\Phi$= 0.4, 0.25, 0.1, and 0 fractions of water are presented in
Figure~\ref{fg4} by the four dashed lines counted from top to bottom,
respectively. By comparing the dashed lines in figures.~\ref{fg2}--\ref{fg4},
one can conclude that decreasing static dielectric permittivity of a
semiconductor substrate results in an increased impact of the fraction
of water in the film on the fluctuation-induced pressure.

\subsection{Comparison between different  substrates}

\begin{figure}[!b]
\vspace*{-3.5cm}
\centerline{\hspace*{0.cm}
\includegraphics[width=3.9in]{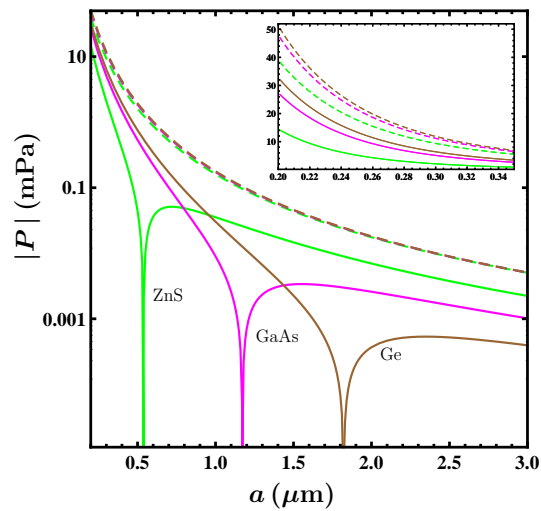}}
\vspace*{-3.1cm}
\caption{\label{fg5}
The magnitudes of the Casimir pressure in peptide film
containing $\Phi$= 0.25 fraction of water deposited on ZnS, GaAs,
and Ge substrates in the dielectric state are shown as the functions
of film thickness by the solid lines. For the semiconductors in
metallic state, the pressure magnitudes are shown by the three dashed
lines counted from top to bottom for Ge, GaAs, and ZnS substrates,
respectively. The region of small film thicknesses is shown in the
inset on an enlarged scale. }
\end{figure}

To illustrate in a more illuminative manner the difference in the
fluctuation-induced pressures in some fixed peptide film deposited
on different semiconductor substrates, we consider the film containing
the 25\% fraction of water. In Figure~\ref{fg5}, the magnitudes of the
Casimir pressure in this film deposited on ZnS, GaAs, and Ge substrates
in the dielectric state are shown by the solid lines as the functions
of film thickness. The three dashed lines counted from top to bottom
present the Casimir pressure in the same peptide film deposited on
Ge, GaAs, and ZnS substrates in the metallic state, respectively.
In the inset, the region of small film thicknesses is shown on an
enlarged scale allowing clear discrimination among the dashed lines.

For the dielectric-type semiconductors, the Casimir pressure in the
peptide film vanishes for the film thicknesses of 0.537, 1.172, and
1.824 $\mu$m for ZnS, GaAs, and Ge substrates, respectively. This is
in agreement with Figures~\ref{fg2}--\ref{fg4} and illustrates the
dependence of the pressure roots on the static dielectric permittivity
of semiconductor materials mentioned above.

In the case of metallic-type semiconductor substrates, a disposition
of the dashed lines in the inset to Figure~\ref{fg5} can also be
connected with a relationship between the static dielectric
permittivities of these substrates. Thus, the top dashed line and that
one below it are rather close. They are for Ge and GaAs substrates
with $\epsilon^{(s)}= 16.2$ and 13.0, respectively. As to the bottom
dashed line related to a ZnS substrate, it is more distant from the
first two in agreement with much lower value of the static permittivity
of ZnS, $\epsilon^{(s)}(0)= 8.35$.

\section{Discussion}

In this paper, we have considered the fluctuation-induced pressure
in peptide films deposited on substrates made of widely employed
semiconductors GaAs, Ge, and ZnS. The dielectric permittivities for
peptide films containing different fractions of water at the
Matsubara frequencies found in the previous literature were used
in computations. The fluctuation-induced (Casimir) pressure in
peptide films was investigated in the framework of fundamental
Lifshitz theory for both dielectric- and metallic-type semiconductor
substrates. The dielectric properties of these substrates were
determined using the available optical data for the complex indices
of refraction of GaAs and Ge and the analytic representation for ZnS.
The electric conductivity of substrates was omitted for the
dielectric-type semiconductors and included for semiconductors in
the metallic state, i.e., using the experimentally and
thermodynamically consistent approach to the Lifshitz theory
\cite{6,43,47,50,51,52,53,54,55,56,57,58}.

According to the results obtained, the Casimir pressure in peptide
films deposited on dielectric-type semiconductor substrates is
repulsive for sufficiently thin films, vanishes for the films of
definite thickness and becomes attractive for thicker films in
the most of cases. The value of film thickness ensuring the null
Casimir pressure increases with decreasing fraction of water in the
film and with increasing static dielectric permittivity of
substrate materials. For the metallic-type semiconductor substrates,
the Casimir pressure in peptide coatings is always repulsive. It
decreases with decreasing fraction of water in the film and with
decreasing static dielectric permittivity of a semiconductor
substrate defined in the dielectric state.

\section{Conclusions}

Peptide films do not possess the crystal structure and their properties are
significantly different from the properties of metallic and dielectric ones.
They consist of big asymmetric molecules and their dielectric permittivities
are not yet investigated over sufficiently wide frequency regions.

Taking into account that peptide coatings are used in organic
electronics, the problem of their stability with further decrease
of the characteristic sizes of microdevices and film thicknesses
may become important not only for fundamental science but also for
its technological applications. The obtained results can be used
as an indication of the ranges of film and substrate parameters
which lead to the attractive Casimir pressure in the film and,
thus, are favorable for its stability.

\funding{The work of O.Yu.T. was supported by the
Russian Science Foundation under Grant No.\ 21-72-20029. G.L.K.\ and
V.M.M.\ were partially supported by the Peter the Great Saint
Petersburg Polytechnic University in the framework of the Russian state
assignment for basic research (Project No.\ FSEG-2020-0024). The work
of V.M.M.\ was also supported by the Kazan Federal University
Strategic Academic Leadership Program.
}

\end{paracol}
\reftitle{References}

\end{document}